\documentclass{PoS}

\usepackage{amsmath}

\title{Theory News Higgs}

\ShortTitle{Higgs Theory News}

\author{\speaker{Joachim Brod}\\%\thanks{A footnote may follow.}\\
        Fakult\"at f\"ur Physik, TU Dortmund, D-44221 Dortmund, Germany\\
        E-mail: \email{joachim.brod@tu-dortmund.de}}

\abstract{I review new developments in Higgs physics, with a focus on
  Yukawa couplings in and beyond the standard model. In particular, I
  discuss different methods of measuring the light Yukawas, new
  sources of CP violation in the Higgs sector, and lepton flavor
  violation. }

\FullConference{16th International Conference on B-Physics at Frontier Machines\\
		2-6 May 2016\\
		Marseille, France}

\begin{document}

\section{Introduction}

The Higgs boson plays a central role in the standard model (SM): On
the one hand, the Higgs mechanism spontaneously breaks the electroweak
gauge symmetry, providing mass terms for the $W$ and $Z$ bosons and
unitarizing gauge boson scattering amplitudes at energies beyond the
electroweak scale. On the other hand, it sets the scale for the
fermion masses via the Yukawa interactions.

With the measurement of the Higgs mass, all SM parameters are fixed;
in particular, all Higgs couplings are determined. This opens a window
to new physics (NP) via the precision measurement of these
couplings. For instance, the SM Yukawa couplings are real and
diagonal; any deviation would be a clear sign of NP. 

While the Higgs couplings to heavy particles seem to agree reasonably
well with SM predictions~\cite{Khachatryan:2016vau}, we know much less
experimentally about the couplings to the light fermions. A simple
understanding of how we can change the Yukawa couplings can be
obtained by parameterizing NP contributions in terms of
higher-dimensional operators. The terms $y_t (\bar Q_{L} t_{R} H^c) +
\text{h.c.}$, for instance, lead to the top-quark mass $ m_t =
\frac{y_t v}{\sqrt{2}}$ and the corresponding top-Higgs Yukawa
coupling $\frac{y_t}{\sqrt{2}}$. Consider now the contribution of a
dimension-six effective operator of the form $\frac{H^\dagger
  H}{\Lambda^2} (\bar Q_{L} t_{R} H^c) + \text{h.c.}$ to the SM
Lagrangian. (You could think of the scale $\Lambda$ as the mass of a
heavy vector-like fermion that has been integrated out.) This operator
will lead to additional contributions to the top-quark mass $\delta
m_t \propto \frac{(v/\sqrt{2})^3}{\Lambda^2}$ and Yukawa coupling
$\delta y_t \propto 3 \frac{(v/\sqrt{2})^2}{\Lambda^2}$. The relative
factor of $3$ between the two new contributions breaks the strict
alignment of mass terms and Yukawa couplings in the SM. This will lead
to off-diagonal and, potentially, imaginary contributions to the
Yukawa couplings after rotating to the mass eigenstates, signalling
the presence of flavor-changing neutral currents (FCNC) and CP
violation.

\section{Size of Yukawa Couplings}

I will first discuss the prospects of measuring the Yukawa couplings,
in particular, to the light fermions. Are there NP models that can
lead to substantial deviations in the Yukawa couplings? One model that
can enhance the light Yukawas by factors of order ten was given by
Giudice and Lebedev~\cite{Giudice:2008uua}. While the simplest version
of this model is already excluded by the measurement of the partial
decay width of the Higgs into bottom quarks, a modified version
employing a Two-Higgs-doublet model is still
viable~\cite{Bishara:2015cha} (see also~\cite{Bauer:2015fxa}). For the
opposite scenario with vanishing light Yukawa couplings see,
e.g.,~\cite{Ghosh:2015gpa}.

Measuring the Yukawa couplings to light fermions with processes
involving off-shell Higgs bosons is extremely difficult, as the
neutral Higgs current always competes with the much larger neutral
currents induced by gluon, photon, or $Z$-boson exchange. One
possibility to circumvent these difficulties is to study the decay of
on-shell Higgs bosons into a photon and a vector meson ($\phi$,
$J/\Psi$, $\Upsilon$)~\cite{Isidori:2013cla, Bodwin:2014bpa,
  Kagan:2014ila, Koenig:2015pha}. The interference of the diagrams
where the Higgs directly couples to the light quark currents and those
where the Higgs decays into a photon and an off-shell photon or $Z$,
converting into the vector meson, leads to sensitivity to the
corresponding Yukawa couplings ($s$, $c$, $b$). The branching ratios
turn out to be very small. They are of the order of $10^{-6}$ for
$h\to\phi\gamma$ and $h\to J/\Psi\gamma$, and, maybe somewhat
surprisingly, of the order of $10^{-9}$ for $h\to\Upsilon\gamma$. In the
latter case the two amplitudes accidentally cancel almost completely,
leading to an increased sensitivity for deviations in the bottom
Yukawa. Unfortunately, the small branching ratios make these processes
very difficult to observe at the LHC.

There are several ways to constrain the charm Yukawa
coupling~\cite{Perez:2015aoa}. Apart from exclusive Higgs decays
discussed above, information can be obtained from the measurement of
the total Higgs decay width via the invariant mass distribution in the
$h\to4\ell$ and $h\to\gamma\gamma$ channels, yielding
$|\kappa_c|\lesssim 120 - 150$. Furthermore, heavy-flavor tagging
currently gives a bound of the order of $|\kappa_c|\lesssim 230$,
while future improvement in charm tagging can tighten thayt
constraint~\cite{Perez:2015lra}. Finally, combining all Higgs data in
a global fit leads to the currently strongest constraint
$|\kappa_c|\lesssim 6.2$~\cite{Perez:2015aoa}.

What do we know about the electron Yukawa? In fact, the best current
bound on the absolute value obtains from direct searches for $h\to
e^+e^-$ at the LHC~\cite{Khachatryan:2014aep}, leading to
$|\kappa_e|<611$~\cite{Altmannshofer:2015qra} (we use the notation
$\kappa_f \equiv y_f/y_f^\text{SM}$ for the Yukawa coupling of any
fermion $f$ in terms of its SM value). This bound is expected to go
down to $|\kappa_e|\lesssim 150$ at the 14\,TeV LHC with 3000/fb of
data, and to roughly $|\kappa_e|\lesssim 75$ at a future 100\,TeV
collider with the same amount of data. A future $e^+e^-$ collider,
collecting 100/fb on the Higgs resonance, would be sensitive to
$|\kappa_e|\sim 15$~\cite{Altmannshofer:2015qra}\footnote{During this
  conference, it was pointed out to me that even SM sensitivity could
  be achieved at FCCee [Stephane Monteil, private communication].}.

It is interesting to compare these bounds to limits obtained from
indirect probes. The anomalous magnetic moment of the electron
$(g-2)_e$ is, via Barr-Zee-type diagrams, proportional to the electron
Yukawa coupling. Usually $(g-2)_e$ is used to define the
fine-structure constant $\alpha$; however, given an independent
determination of $\alpha$, the MDM is a sensitive probe of
NP~\cite{Giudice:2012ms}. Using the measurements of
$\alpha$~\cite{Bouchendira:2010es} and $(g-2)_e$~\cite{Hanneke:2010au}
yields a bound $|\kappa_e|\lesssim
3000$~\cite{Altmannshofer:2015qra}. This is weaker than the current
LHC bound; however, it depends linearly on the electron Yukawa and the
sensitivity is expected to increase by a factor of ten in the next few
years~\cite{Giudice:2012ms}.

\section{CP Violation in the Higgs Sector}

For successful baryogenesis new sources of CP violation are
needed~(see, e.g.,~\cite{Bernreuther:2002uj}). They could be provided
by complex Yukawa couplings; see~\cite{Huber:2006ri} for a minimal
setup. Can we test this scenario?

Modifying the top Yukawa will change Higgs production via gluon fusion
and the $h\to\gamma\gamma$ decay, while changing the bottom Yukawa
will change all Higgs branching ratios. LHC measurements currently
still allow for order one deviations from SM predictions in the Higgs
couplings to the third generation. However, electric dipole moments
(EDMs) induced via Barr-Zee diagrams~\cite{Weinberg:1989dx,
  Barr:1990vd} tend to yield much stronger constraints. For instance,
the recent measurement of the electron EDM, $d_e/e < 8.7\times
10^{-29}\,$cm~\cite{Baron:2013eja} leads to Im$\,\kappa_t\lesssim
0.01$~\cite{Brod:2013cka}. (See~\cite{Jung:2013hka, Chien:2015xha} for
a comprehensive discussion of theory uncertainties.) For the bottom
and $\tau$, EDMs lead to constraints comparable to those from
LHC~\cite{Brod:2013cka}. Information from differential decay
distributions can, however, provide additional information on the CP
phase at colliders~\cite{Harnik:2013aja, Galanti:2015pqa}.

A complete analytic results for the two-loop electron-Yukawa
contributions to the electron EDM has only been given
recently~\cite{Altmannshofer:2015qra} and leads to
Im$\,\kappa_e\lesssim 0.017$ (using the same measurement as above). A
corresponding full result for the light-quark Yukawas has not yet been
published; a preliminary calculation, however, yields
Im$\,\kappa_u\lesssim 0.08$ and Im$\,\kappa_d\lesssim 0.02$ for the
up- and down-quark Yukawa, respectively~\cite{BWIP}. (Here, I used the
most recent measurement of the neutron EDM from~\cite{Baker:2006ts}).

\section{Lepton Flavor Violation}

Some excitement was caused recently by a hint for a non-zero branching
ratio of the lepton-flavor violating decay
$h\to\tau\mu$~\cite{Khachatryan:2015kon}. Indeed, precision
observables in the lepton sector (for instance, $(g-2)_\mu$, EDMs,
$\tau\to 3\mu$, $\mu\to 3e$, $\tau \to \mu \gamma$, $\mu \to e\gamma$)
allow for Br$(h\to\tau\mu)={\cal O}(10\%)$~\cite{Harnik:2012pb}. The
most ``directly related'' precision observable is the rare decay
$\tau\to\mu\pi\pi$~\cite{Celis:2013xja}. While the CMS result implies
$\text{BR}(\tau \to \mu \pi^+ \pi^-) < 1.6 \times 10^{-11}$, the
current bounds are $\text{BR}(\tau \to \mu \pi^+ \pi^-) \lesssim
\text{few} \times 10^{-8}$ from Belle~\cite{Miyazaki:2012mx} and
$\text{BR}(\tau \to \mu \pi^0 \pi^0) < 1.4 \times 10^{-5}$ from
Cleo~\cite{Bonvicini:1997bw}. Stronger constraints could are expected
at Belle~II.

The CMS measurement of
Br$(h\to\tau\mu)=(0.84^{+0.39}_{-0.37})\%$\footnote{Note that the
  significance of this measurement decreased after analysis of new
  data~\cite{CMS:2016qvi}.} corresponds to an average flavor-changing
$\mu\tau$ Yukawa of $\sqrt{|Y_{\tau\mu}|^2 + |Y_{\mu\tau}|^2} = (2.6
\pm 0.6) \times 10^{-3}$~\cite{Altmannshofer:2015esa}. In general, a
(large) $h\to\tau\mu$ branching ratio will imply a large $\tau\mu$
dipole operator, as at least on of the particles in the loop has to be
electrically charged. To be consistent with precision constraints,
this then requires either fine tuning, or a second source of
electroweak symmetry breaking~\cite{Altmannshofer:2015esa}. A plethora
of models has been proposed (e.g., 2HDM~\cite{Crivellin:2015mga,
  Altmannshofer:2016oaq, Bizot:2015qqo};
leptoquarks~\cite{Cheung:2015yga, Dorsner:2015mja}; new strong
interactions~\cite{Altmannshofer:2015esa}). Interestingly, it is not
possible to explain a large $h\to\tau\mu$ branching ratio within the
MSSM: all (necessarily fine-tuned) solutions consistent with precision
bounds are ruled out by the existence of charge-breaking
vacua~\cite{Aloni:2015wvn}.

\section{New Ideas}

At last, I would like to mention two new ideas how to constrain the
light-quark Yukawa couplings. 

The first is to constrain the product of electron and light-quark
quark Yukawa via so-called ``atomic clock
transitions''~\cite{Delaunay:2016brc}. The point-like and attractive
Higgs force will induce small changes in the characteristic energy
(frequency) differences in suitable atomic transitions. Since the
Higgs contribution cannot be switched off, the measurement of
transitions in several isotopes is required; measuring the
``difference of differences'' allow for the elimiation of hadronic
uncertainties. In this way, an independent measurement of the
light-fermion Yukawas could be possible, with a sensitivity comparable
to that of the current LHC bounds or better~\cite{Delaunay:2016brc}.

The second idea is to constrain the light Yukawas by measuring the
charge asymmetry in the process $hW^\pm \to (\ell^\pm)(\ell^\pm\nu
jj)$~\cite{YWIP}. It encodes mainly the underlying pdf
asymmetry. While the dominant SM contribution is the radiation of a
higgs boson off a $W$ boson (corresponding to a charge asymmetry of
order $25\%$), for enhanced light Yukawas, the emission of a higgs
from an initial light-quark line becomes comparable, leading to a
sensitivity to these couplings. Additional contributions to the
asymmetry between $-30\%$ and $+5\%$ can be expected~\cite{YWIP}.

\section{Summary}

The Higgs couplings are completely determined in the SM; that is why
we need to measure them! Any deviation (for instance, CP-violating or
flavor-changing Yukawa couplings) would be a clear sign of NP. The
measurement of the light-fermion Yukawa couplings, in particular, is a
difficult experimental problem, and an interesting interplay between
collider observables and precision probes exists. The discovery of the
Higgs boson opened a new window to search for new physics in the Higgs
sector, which quickly became an active and exciting new field of
research.

\end{document}